# A generalized Hamiltonian formulation of the principle of virtual work

D. H. Delphenich ([1])
Spring Valley, OH 45430

**Abstract:** The author's previous derivation of a variational principle from the total work functional, as a generalization of the first variation of an action functional, is extended by deriving a corresponding generalization of the Hamiltonian formulation of that action functional. Some consequences of it are that one can decompose the Lie brackets of arbitrary vector fields on symplectic mechanics into a sum of terms that involve the Poisson brackets of the functions that appear in the normal form of the Pfaffian that is symplectic-dual to the vector field and that one can also generalize the Hamilton-Jacobi equation to a system of nonlinear first-order partial differential equations for the contact field that one uses in order to obtain them.

**Introduction.** – The basic problem that will be addressed in this article is the problem of finding first principles that will lead to equations of motion for physical systems as a consequence. Such first principles exist in varying degrees of generality in terms of the physical systems that they are capable of describing. Typically, the physical systems will include constraints in one form or another.

Perhaps the most general first principle is Gauss's principle of least constraint ([2]), which includes the possibility that the constraints that apply to the motion are defined by inequalities, and not equalities. For instance, the example that was given by Gibbs [**2**] was that of a marble rolling down the surface of an inverted bowl in the presence of gravity. Depending upon the initial velocity of the marble, the trajectory could either be on that surface, which would be an equality (or "two-sided") constraint, or it could leave the surface and follow a ballistic trajectory in the space above the surface, which would be an inequality (or "one-sided") constraint.

If one restricts oneself to equality (or two-sided) constraints, which do not have to be perfect or holonomic, and subjects the motion to applied forces that do not have to be conservative then the next level of generality is defined by the principle of virtual work, in conjunction with d'Alembert's principle ([3]). That first principle leads to the Lagrange equations in their first form. Although it is commonly thought that despite the interpretation of virtual displacements as constrained variations, the principle of virtual work is not manifestly a variational principle, as we shall see, it can be used as the basis for one.

---

([1]) E-mail: feedback@neo-classical-physics.info, Website: neo-classical-physics.info
([2]) The author has compiled a collection of his own translations of various papers on that topic into a book [**1**] that is available at this website as a free PDF download.
([3]) For the most directly relevant textbooks on the classical results of analytical mechanics, one might consult any of the references cited in [**3**].

When one deals with only conservative forces and perfect holonomic constraints one will get to the Lagrange equations in their second form. They take the form of the Euler-Lagrange equations for a variational principle that is commonly referred to as the *least action principle* or *Hamilton's principle* ([1]). In any event, that means that not all physical systems in nature can be derived from a least-action principle.

In some previous work by the author [**4**], it was shown that one could derive equations of motion that are more general than the Euler-Lagrange equations by starting from essentially the first variation functional as something akin to a "total" virtual work functional. Thus, the level of generality of the variational principle is essentially equivalent to the principle of virtual work, combined with d'Alembert's principle, and includes Lagrange's equations in their first form.

Many modern physicists find that the Hamiltonian formulation of the least-action principle and the resulting equations of motion are often more convenient than the Lagrangian formulation. This is especially true when one gets into mathematical optics and the quantum wave mechanics of Louis de Broglie and Erwin Schrödinger, which was based in the optical-mechanical analogy that essentially goes back to Maupertuis and Fermat [**5**]. Thus, since the level of generality of the Hamiltonian methodology is roughly equivalent (up to the invertibility of the "momentum map," which is usually the case) to the Lagrangian methodology, there is some question regarding whether the Hamiltonian formalism can be generalized in a manner that would be analogous to the aforementioned generalization of the least-action principle to a total virtual work principle. That is then the expressed goal of the present work.

Section 1 is a review of the previous approach to starting the calculus of variations with the first-variation functional, rather than the action functional, and some of the results that it implies. Section 2 then addresses how one might generalize the Legendre transformation by generalizing essential the differential *dH* of the Hamiltonian. The third section then shows how to derive generalized Hamilton equations from the results of the previous section. Section 4 then addresses how one might generalize the Poisson brackets on a symplectic manifold to the Lie brackets of arbitrary vector fields and obtains a decomposition of the latter that includes the Poisson brackets of the functions that occur in the Pfaffian normal forms for the 1-forms that are symplectic dual to the vector fields. Finally, the Hamilton-Jacobi equation is generalized accordingly. The article then concludes with a discussion of the possible applications of the result to established topics in Hamiltonian physics. In each section, the conventional derivation of the Lagrangian and Hamiltonian ideas will be first present, and then the generalization of that formalism to the total virtual work functional will be given.

Although repeated reference will be made to the theory of Pfaff problem ([2]), in fact, the only essential idea that will be used is that a Pfaffian (i.e., 1-form) can be expressed in one of two normal forms. The choice that we shall make is the one that naturally extends an exact 1-form to an exact 1-form plus inexact corrections.

---

  ([1])  Although the further that one goes back in time, the more that mathematicians and physicists object to the precision of that terminology!
  ([2])  The author has produced a survey article [**6**] on the theory of the Pfaff problem and how it applied to various topics in physics.

**1. Review of previous results.** – Only the essential definitions and results from the previous articles by the author will be summarized here, for the sake of completeness.

*a. Conventional derivation of equations of motion from the least-action principle.* – In order to define a (first-order) action functional on curve segments in an $n$-dimensional configuration manifold $M$, one first defines the manifold of first-order kinematical states of moving points in $M$. That will take the form of the manifold $J^1(\mathbb{R}, M)$ of *1-jets of differentiable curves in M*. A *1-jet* $j_x^1 \gamma$ of a differentiable curve $\gamma(t)$ through a point $x \in M$ is the equivalence class of all other such curves through $x$ that have the same tangent vector (i.e., velocity) in $T_xM$. Note that the definition of a 1-jet of differentiable curves through $x \in M$ is identical to the definition of a tangent vector in $T_xM$.

Let $(U, x^i)$ be a local coordinate chart on $M$ about $x$, where $U$ is an open neighborhood of $x$ and $x^i : U \to \mathbb{R}^n, p \mapsto x^i(p)$ is a set of $n$ coordinate functions on $U$ that define a homeomorphism of $U$ with $\mathbb{R}^n$. The 1-jet $j_x^1 \gamma$ can then be described by the set of numbers $(t, x^i, v^i)$, where $v^i$ are the components of the tangent vector in $T_xM$ that all of the curves are tangent to with respect to the natural frame field $\{\partial/\partial x^i, i = 1, \ldots, n\}$ that is associated with the coordinate chart. Thus, the set of numbers $(t, x^i, v^i)$ will serve as a local coordinate chart on $J^1(\mathbb{R}, M)$ about the point $j_x^1 \gamma$. Once again, the only difference between a coordinate chart on the subset $T(U)$ of the tangent bundle $T(M)$ that lies over $U$ and a coordinate chart about $j_x^1 \gamma$ in $J^1(\mathbb{R}, M)$ is the addition of the curve parameter $t$. Since we shall be mostly doing local calculations in what follows, that definition of a coordinate chart on $J^1(\mathbb{R}, M)$ will suffice for our purposes.

Now, there is a fundamental difference between a differentiable curve $x(t)$ in $M$ and a curve in $J^1(\mathbb{R}, M)$, which will take the local form $(t, x(t), v(t))$, since in general the tangent vector $(t, x^i(t), v^i(t))$ in $T_{x(t)}M$ does not have to be the actual velocity vector:

$$(1.1) \qquad \dot{x}^i(t) = \left.\frac{dx^i}{dt}\right|_t$$

to the curve at the point $x(t)$. Indeed, the case in which one actually has:

$$(1.2) \qquad v^i(t) = \dot{x}^i(t)$$

will be regarded as the *integrable* case of a curve in $J^1(\mathbb{R}, M)$, and if $x^i(t)$ is a differentiable curve in $M$ then the curve in $J^1(\mathbb{R}, M)$ that it defines by differentiation:

$$(1.3) \qquad j^1 x(t) = (t, x^i(t), \dot{x}^i(t))$$

will be called the *1-jet prolongation* of $x^i(t)$, and a curve in $J^1(\mathbb{R}, M)$ for which one has (1.2) will be called *integrable*. One can envision the difference between an integrable curve and a non-integrable one by imagining that in the integrable case, the tangent vector points in the direction of motion along the curve at each point, while in the non-integrable case, it can point in any direction.

An example of a non-integrable curve is given by the way that the curve looks in a rotating frame field with an angular velocity that is described by a time-varying skew-symmetric matrix $\omega^i_j(t)$, namely:

$$(1.4) \qquad v^i(t) = \dot{x}^i(t) + \omega^i_j(t)\, x^i(t).$$

A *variation* of the curve $x^i(t)$ in $M$ will be a vector field:

$$(1.5) \qquad \delta x(t) = \delta x^i(t)\frac{\partial}{\partial x^i}$$

along the curve that will typically not be integrable. Indeed, since it is regarded as the infinitesimal generator of a one-parameter family of deformations of the curve $x^i(t)$, if it were tangent (in the integrable sense) then that would represent a deformation of the curve parameterization. Thus, the variations that will be used in what follows will typically be *transverse* to the curve (i.e., not tangent in the integrable sense).

One can also define variations of curves in $J^1(\mathbb{R}, M)$, namely, if the curve takes the local form $s(t) = (t, x(t), v(t))$ then a variation of it will take the general form:

$$(1.6) \qquad \delta s(t) = \delta t(t)\frac{\partial}{\partial t} + \delta x^i(t)\frac{\partial}{\partial x^i} + \delta v^i(t)\frac{\partial}{\partial v^i}.$$

Typically, we shall restrict ourselves to variations for which $\delta t(t) = 0$, for the reasons that were just stated.

Once again, one can define a variation of a curve in $J^1(\mathbb{R}, M)$ to be *integrable* iff:

$$(1.7) \qquad \delta v^i(t) = \left.\frac{d\,\delta x^i}{dt}\right|_t.$$

We are now in a position to define an action functional that associates a number $S[x(t)]$ with every differentiable curve segment $x(t)$ in $M$ that goes from $x_0 = x(t_0)$ to $x_1 = x(t_1)$. We first define a *Lagrangian function* (or simply *Lagrangian*), which is a twice-continuously-differentiable (i.e., $C^2$) function on $J^1(\mathbb{R}, M)$:

$$(1.8) \qquad L = L(t, x, v).$$

We next "prolong" the curve $x(t)$ in $M$ to a curve $j^1 x(t) = (t, x(t), \dot{x}^i(t))$ in $J^1(\mathbb{R}, M)$, so the function:

(1.9) $$L(j^1 x(t)) = L(t, x(t), \dot{x}^i(t))$$

will become simply a function of $t \in \mathbb{R}$. When it is multiplied by the 1-form $dt$ on $\mathbb{R}$, we will get a 1-form $L(j^1 x(t))\, dt$ on $\mathbb{R}$ that can (presumably) be integrated from $t_0$ to $t_1$. That integral:

(1.10) $$S[x(t)] = \int_{t_0}^{t_1} L(j^1 x(t))\, dt = \int_{t_0}^{t_1} L(t, x^i(t), \dot{x}^i(t))\, dt$$

will be what we call the *action functional* for differentiable curve segments in $\mathbb{R}$ (relative to $L$).

One can, for the sake of heuristics, imagine that if the curve segments in $M$ define an "infinite-dimensional differentiable manifold" (which they do not, unless one adds a lot of analytical overhead) then $S$ will be essentially a "differentiable function" on that manifold. The gist of the calculus of variations can then be regarded as "the calculus of infinity variables." In particular, one looks for the "critical points" of that action functional, which will be the curve segments for which its "differential" vanishes.

In order to make that more rigorous without going into the analytical details – i.e., in order to define the *calculus* of variations, and not the *analysis* of variations – We shall define the *first variation* $\delta S$ of the action functional, which will play the role of the differential of the functional $S$. The *first variation* of the action functional $S[\ldots]$ is a linear functional on vector fields along curve segments that one defines by saying that if $\mathbf{X}(t)$ is a vector field along the curve segment $s(t)$ in $J^1(\mathbb{R}, M)$, as defined above, then the value of $\delta S[\mathbf{X}(t)]$ will be obtained from:

(1.11) $$\delta S[\mathbf{X}(t)] = \int_{t_0}^{t_1} dL(\mathbf{X}(t))\, dt,$$

in which we have abbreviated the notation for clarity. In particular, when $\mathbf{X}(t)$ takes the form of a variation $\delta s(t)$, we shall use the notation:

(1.12) $$\delta L(t) = dL(\delta s(t)).$$

Now, in general, one will have:

(1.13) $$dL = \frac{\partial L}{\partial t} dt + \frac{\partial L}{\partial x^i} dx^i + \frac{\partial L}{\partial v^i} dv^i,$$

and if one defines a transverse variation of $s(t)$ by:

(1.14) $$\delta s(t) = \delta x^i(t) \frac{\partial}{\partial x^i} + \delta v^i(t) \frac{\partial}{\partial v^i}$$

then one will have:

(1.15) $$\delta L(t) = \frac{\partial L}{\partial x^i} \delta x^i + \frac{\partial L}{\partial v^i} \delta v^i.$$

If one now assumes that $\delta L$ is integrable, so:

(1.16) $$\delta L(t) = \frac{\partial L}{\partial x^i} \delta x^i + \frac{\partial L}{\partial v^i} \frac{d}{dt} \delta x^i,$$

then with an application of the product rule for differentiation (i.e., integration by parts), that can be given the form:

(1.17) $$\delta L(t) = \left( \frac{\partial L}{\partial x^i} - \frac{d}{dt} \frac{\partial L}{\partial v^i} \right) \delta x^i + \frac{d}{dt} \left( \frac{\partial L}{\partial v^i} \delta x^i \right),$$

and the first variation of $S[\ldots]$ can be given the form:

(1.18) $$\delta S[j^1 \delta x(t)] = \int_{t_0}^{t_1} \left( \frac{\partial L}{\partial x^i} - \frac{d}{dt} \frac{\partial L}{\partial v^i} \right) \delta x^i \, dt + \left[ \frac{\partial L}{\partial v^i} \delta x^i \right]_{t_0}^{t_1}.$$

If one introduces the notation:

(1.19) $$\frac{\delta L}{\delta x^i} = \frac{d}{dt} \frac{\partial L}{\partial v^i} - \frac{\partial L}{\partial x^i},$$

which shall be called the *variational derivative* of $L$ with respect to $x^i$, then the last equation can be put into the form ([1]):

(1.20) $$\delta S[\delta x(t)] = - \int_{t_0}^{t_1} \frac{\delta L}{\delta x^i} \delta x^i \, dt + \left[ \frac{\partial L}{\partial v^i} \delta x^i \right]_{t_0}^{t_1}.$$

The *principle of least action* in mechanics says that of all curve segments $x(t)$ between two points $x_0$ and $x_1$ in $M$, the "natural" path will be the one for which the action functional $S[x(t)]$ is a minimum. A necessary condition for that is that it should be an extremum, which means that its first variation should vanish for all variations $\delta x(t)$ that satisfy certain boundary (i.e., endpoint) conditions. Typically, those constraints on $\delta x(t)$ are intended to make the bracketed term in the last equation vanish. The two most common are the *fixed endpoint* constraint, which makes the variation $\delta x(t)$ vanish at the endpoints, and the *transversality* condition, which makes the entire term in the brackets vanish at the endpoints. Of course, the most general constraint would be to simply demand that the entire bracket must collectively vanish, but that condition does not seem to get very much attention.

---

([1]) The introduction of the negative sign, which is implicit in the definition of the variational derivative, is to anticipate the fact that in most examples (such as geodesics), one wishes that the acceleration should have a positive sign, and the acceleration usually comes from the time derivative in the variational derivative.

If a curve segment $x(t)$ is an *extremal*, in the sense of a path of least action, then that will mean that $\delta S[\delta x(t)]$ must vanish for every variation $\delta x(t)$ that satisfies suitable constraints that make the bracketed term vanish, and a necessary and sufficient condition for that is that:

$$(1.21) \qquad 0 = \frac{\delta L}{\delta x^i} = \frac{d}{dt}\frac{\partial L}{\partial v^i} - \frac{\partial L}{\partial x^i}.$$

Those are the *Euler-Lagrange* equations of motion. They describe a broad class of mechanical systems in theoretical mechanics, but not, by any means, all of them. In effect, the main limitations usually take the form of saying that one must be dealing with conservative forces and perfect holonomic constraints. Since that does not exhaust the possibilities in physics (viz., natural motions), one might wish to look for a more general way of obtaining more general equations of motion.

One can derive the balance of energy from the Euler-Lagrange equations by first computing the total derivative of $L$ with respect to $t$ along a curve $x(t)$:

$$(1.22) \qquad \frac{dL}{dt} = \frac{\partial L}{\partial t} + \frac{\partial L}{\partial x^i}\frac{dx^i}{dt} + \frac{\partial L}{\partial v^i}\frac{dv^i}{dt}.$$

Substituting the value of $\partial L/\partial x^i$ that is derived from the Euler-Lagrange equations will make that take the form:

$$(1.23) \qquad \frac{dL}{dt} = \frac{\partial L}{\partial t} + \left(\frac{d}{dt}\frac{\partial L}{\partial v^i}\right)\frac{dx^i}{dt} + \frac{\partial L}{\partial v^i}\frac{d}{dt}\frac{dx^i}{dt} = \frac{\partial L}{\partial t} + \frac{d}{dt}\left(\frac{\partial L}{\partial v^i}\frac{dx^i}{dt}\right).$$

If one defines the *total energy* of the motion to be:

$$(1.24) \qquad E = \frac{\partial L}{\partial v^i}\frac{dx^i}{dt} - L$$

then equation (1.23) will imply that:

$$(1.25) \qquad \frac{dE}{dt} = -\frac{\partial L}{\partial t}.$$

Thus, the total energy of motion will be constant along an extremal iff the Lagrangian is time-invariant.

*b. Variational formulation of the principle of virtual work.* – Since the equations of motion, as a system of ordinary differential equations, are mostly due to the vanishing of the first variation functional for all variations of a curve that satisfy the endpoint conditions that one imposes, and the first variation of the action functional is based upon an exact 1-form:

$$(1.26) \qquad dL = \frac{\partial L}{\partial t} dt + \frac{\partial L}{\partial x^i} dx^i + \frac{\partial L}{\partial v^i} dv^i,$$

it would be natural to generalize that 1-form on $J^1(\mathbb{R}, M)$ to something that is not exact, especially because that is usually a question of the exactness of the force 1-form. Thus, we propose to replace $dL$ with a *fundamental 1-form* on $J^1(\mathbb{R}, M)$:

$$(1.27) \qquad \phi = P\,dt + F_i\,dx^i + p_i\,dv^i,$$

whose components take the form of power, force, and momentum, respectively. In general, their functional dependency will be:

$$(1.28) \qquad P = P(t, x, v), \qquad F_i = F_i(t, x, v), \qquad p_i = p_i(t, x, v),$$

which then amount to "mechanical constitutive laws" for the mechanical system in question.

In the case where an action functional for the motion exists, $\phi$ will then take the form $\phi = dL$, and:

$$(1.29) \qquad P = \frac{\partial L}{\partial t}, \qquad F_i = \frac{\partial L}{\partial x^i}, \qquad p_i = \frac{\partial L}{\partial v^i}.$$

Thus, the momentum components $p_i$ will coincide with the conjugate momenta under $L$.

If a variation of a curve $s(t) = (t, x(t), v(t))$ in $J^1(\mathbb{R}, M)$ takes the form of a vector field along that curve:

$$(1.30) \qquad \delta s(t) = \delta t(t)\frac{\partial}{\partial t} + \delta x^i(t)\frac{\partial}{\partial x^i} + \delta v^i(t)\frac{\partial}{\partial v^i}$$

then the evaluation of the 1-form $\phi$ on the vector field $\delta\gamma$ will give the function on $J^1(\mathbb{R}, M)$:

$$(1.31) \qquad \phi(\delta s) = P\,\delta t + F_i\,\delta x^i + p_i\,\delta v^i.$$

The generalization of the first variation functional along the curve $x(t)$ in $M$ is to pull back the function $\phi(\delta s)$ to the real line by way of the 1-jet prolongation of $x(t)$, which is the curve in $J^1(\mathbb{R}, M)$:

$$(1.32) \qquad j^1 x(t) = (t, x(t), \dot{x}(t)).$$

The pull-back of the function $\phi(\delta\gamma)$ to $\mathbb{R}$ is then:

$$(1.33) \qquad \delta W(t) = j^1 x^* [\phi(\delta s)](t) = P(t)\delta t(t) + F_i(t)\delta x^i(t) + p_i(t)\delta v^i(t),$$

which will make the first variation functional on $\delta\gamma$ take the form:

$$(1.34) \qquad W[\delta s] = \int_{t_0}^{t_1} \delta W(t)\, dt = \int_{t_0}^{t_1} j^1 x^* [\phi(\delta s)](t)\, dt.$$

The use of the symbol $W$ is to suggest that the physical meaning of $\delta W$ is basically the virtual work that is done under the virtual displacement $\delta s$, while $W$ will then become the *total virtual work* done along the curve $x(t)$.

Now, one typically does not vary $t$, since that would amount to a change in the speed of parameterization, and a concomitant change in the kinetic energy, so one usually sets $\delta t = 0$. Furthermore, if one assumes that the remaining variation $\delta s$ is also *integrable*, so:

$$(1.35) \qquad \delta v^i = \frac{d}{dt}\delta x^i,$$

then that will make:

$$(1.36) \qquad \delta s(t) = \delta x^i(t)\frac{\partial}{\partial x^i} + \left(\frac{d}{dt}\delta x^i(t)\right)\frac{\partial}{\partial v^i}$$

and

$$(1.37) \qquad \delta W = F_i \delta x^i + p_i \frac{d}{dt}\delta x^i,$$

in which the functional dependency on $t$ has been suppressed.

With the usual application of the product rule for differentiation (i.e., integration by parts), that will take the form:

$$(1.38) \qquad \delta W = \left(F_i - \frac{dp_i}{dt}\right)\delta x^i + \frac{d}{dt}(p_i \delta x^i),$$

and the total virtual work will take the form:

$$(1.39) \qquad W[\delta x] = \int_{t_0}^{t_1}\left[\left(F_i - \frac{dp_i}{dt}\right)\delta x^i\right] dt + [p_i \delta x^i]_{t_0}^{t_1}.$$

When one imposes the customary endpoint conditions (viz., fixed endpoints or transverse variations), the second term will vanish, and the vanishing of $W[\delta x]$ for all $\delta x^i$ will then give the equations of motion in the form:

(1.40) $$F_i = \frac{dp_i}{dt},$$

which has essentially a Newtonian form.

When $\phi = dL$, that will take the form:

(1.41) $$\frac{\partial L}{\partial x^i} = \frac{d}{dt}\frac{\partial L}{\partial \dot{x}^i},$$

which are the Euler-Lagrange equations (also called the "Lagrange equations in their second form" [**3**]).

When:

(1.42) $$F_i = Q_i + \frac{\partial T}{\partial x^i}, \qquad p_i = \frac{\partial T}{\partial \dot{x}^i},$$

one will get:

(1.43) $$\frac{d}{dt}\frac{\partial T}{\partial \dot{x}^i} - \frac{\partial T}{\partial x^i} = Q_i,$$

which are the Lagrange equations in the first form, which are usually derived from the principal of virtual work, combined with d'Alembert's principle [**3**]

One then sees how starting with the vanishing of the "first variation" of a curve, rather than demanding that the action along the curve should be an extremum, can serve as a more general first principle for the derivation of equations of motion.

**2. Generalized Legendre transformation.** – We shall review the conventional way of defining the Legendre transformation and then propose a reasonable generalization of it.

*a. Conventional definition of the Legendre transformation*. – Customarily, in order to convert a Lagrangian function $L(t, x, v)$ on $J^1(\mathbb{R}, M)$ into a Hamiltonian function $H(t, x, p)$ on a manifold $\mathcal{P}(\mathbb{R}, M)$ that we shall discuss shortly, one first defines a "momentum map" that will associate the velocity components $v^i$ with their conjugate momenta with respect to $L$:

(2.1) $$p_i(t, x, v) = \frac{\partial L}{\partial v^i}.$$

The functions $p_i(t, x, v)$ are then the components of a 1-form on $J^1(\mathbb{R}, M)$, namely:

(2.2) $$p = p_i(t, x, v)\, dv^i,$$

that is seen to be part of the 1-form $dL$.

If one assumes that the system of *n* equations (2.1) can be solved for the *n* components of velocity:

(2.3) $$v^i = v^i(t, x, p)$$

then the implicit function theorem says that the *mass matrix:*

(2.4) $$m_{ij}(t, x, v) = \frac{\partial p_i}{\partial v^j} = \frac{\partial^2 L}{\partial v^i \partial v^j}$$

must be invertible. Indeed, in most cases of interest to physics, it will take the form:

(2.5) $$m_{ij} = m(t)\, g_{ij}(t, x),$$

in which $m(t)$ is the time-varying mass of the point along a trajectory $x^i(t)$, and $g_{ij}(t, x)$ are the components of the spatial metric. For instance, the motion might be constrained by a surface that deforms in time (e.g., floating on the surface of the ocean).

Once that has been established, the Legendre transformation is achieved by defining the Hamiltonian function that corresponds to *L* by way of:

(2.6) $$H(t, x, p) = p_i v^i(t, x, p) - L(t, x, v(p)).$$

Although the manifold $\mathcal{P}(\mathbb{R}, M)$, which we shall call the *phase space of the configuration manifold M*, can be defined in a more rigorous way (cf., e.g., [**7**]), since we shall be mostly concerned with calculations of a local nature here, it will suffice to say that its local coordinate charts take the form $(t, x, p)$. In the case of time-invariant Lagrangians and Hamiltonians, just as the manifold $J^1(\mathbb{R}, M)$ can be reduced to the tangent bundle $T(M)$, similarly, the manifold $\mathcal{P}(\mathbb{R}, M)$, can be associated with the cotangent bundle $T^*M$.

*b. Generalized Legendre transformation.* – Since a Lagrangian function *L* will exist for the fundamental 1-form $\phi$ iff it is exact, in order to generalize the Legendre transformation, one will have to first look at the corresponding 1-form that is obtained from *H* by differentiation:

$$dH = dp_i\, v^i + p_i\, dv^i - dL = dp_i\, v^i + p_i\, dv^i - \frac{\partial L}{\partial t} dt - \frac{\partial L}{\partial x^i} dx^i - \frac{\partial L}{\partial v^i} dv^i,$$

and when one makes the association (2.1), that will make:

(2.7) $$dH = -\frac{\partial L}{\partial t} dt - \frac{\partial L}{\partial x^i} dx^i + v^i dp_i.$$

If one generalizes this 1-form in the same way as before, i.e., replacing $\partial L / \partial t$ with $P$ and $\partial L / \partial x^i$ with $F_i$ then one can define the 1-form that generalizes $dH$ to be:

(2.8) $$\eta = -P\,dt - F_i\,dx^i + v^i\,dp_i,$$

in which one must now have:

(2.9) $$P = P(t, x, p), \qquad F_i = F_i(t, x, p), \qquad v^i = v^i(t, x, p).$$

By abuse of notation, one can say that the relationship between $\phi$ and $\eta$ is:

(2.10) $$\eta = d(p_i v^i) - \phi.$$

If one expresses $\phi$ in its normal form as a Pfaffian:

(2.11) $$\phi = dL + \mu_a\,dv^a \qquad (a = 1, \ldots, p)$$

in which $\mu_a$ and $v^a$ are both functions of $(t, x, v)$ then one can say that:

$$\eta = d(p_i v^i - L) - \mu_a\,dv^a,$$

and upon expressing $v$ as a function of $(t, x, p)$ and defining $H$ as in (2.6), one can express $\eta$ in the Pfaffian normal form:

(2.12) $$\eta = dH - \mu_a\,dv^a.$$

If we expand the differentials in this then that will give:

(2.13) $$\eta = \left(\frac{\partial H}{\partial t} - \mu_a \frac{\partial v^a}{\partial t}\right) dt + \left(\frac{\partial H}{\partial x^i} - \mu_a \frac{\partial v^a}{\partial x^i}\right) dx^i + \left(\frac{\partial H}{\partial p_i} - \mu_a \frac{\partial v^a}{\partial p_i}\right) dp_i.$$

**3. Generalized Hamilton equations.** – If we compare the corresponding components of the 1-form in this last equation with the ones in (2.8) then that will give:

(3.1) $$P = -\frac{\partial H}{\partial t} + \mu_a \frac{\partial v^a}{\partial t}, \qquad F_i = -\frac{\partial H}{\partial x^i} + \mu_a \frac{\partial v^a}{\partial x^i}, \qquad v^i = \frac{\partial H}{\partial p_i} - \mu_a \frac{\partial v^a}{\partial p_i}.$$

If we define the usual symplectic 2-form by:

(3.2) $$\Omega = dp_i \wedge dx^i$$

then a vector field that takes the general form:

(3.3) $$\mathbf{X} = X^t \frac{\partial}{\partial t} + X^i \frac{\partial}{\partial x^i} + X_i \frac{\partial}{\partial p_i},$$

in which all components are functions of $(t, x, p)$, will be associated with a 1-form by way of $\Omega$ and the interior product operator:

(3.4) $$i_\mathbf{X} \Omega = X_i \, dx^i - X^i \, dp_i .$$

Note that the component $X^t$ does not play a role in that expression.

Thus, if the vector field $\mathbf{X}$ takes the form of the velocity vector field for a differentiable curve of the form $(t, x(t), p(t))$, i.e.:

(3.5) $$\mathbf{v}(t) = \frac{dx^i}{dt} \frac{\partial}{\partial x^i} + \frac{dp_i}{dt} \frac{\partial}{\partial p_i},$$

so

(3.6) $$i_\mathbf{v} \Omega = \frac{dp_i}{dt} dx^i - \frac{dx^i}{dt} dp_i$$

a comparison of that with (3.1) will give:

(3.7) $$\frac{dx^i}{dt} = \frac{\partial H}{\partial p_i} - \mu_a \frac{\partial v^a}{\partial p_i}, \qquad \frac{dp_i}{dt} = -\frac{\partial H}{\partial x^i} + \mu_a \frac{\partial v^a}{\partial x^i},$$

which is clearly a generalization of the usual canonical equations that Hamilton deduced, except for the one that relates to the total time derivative of $H$. That equation can be deduced by direct calculation:

$$\frac{dH}{dt} = \frac{\partial H}{\partial t} + \frac{\partial H}{\partial x^i} \frac{dx^i}{dt} + \frac{\partial H}{\partial p_i} \frac{dp_i}{dt},$$

and with the substitutions in (3.7), that will take the form:

$$\frac{dH}{dt} = \frac{\partial H}{\partial t} + \frac{\partial H}{\partial x^i} \left( \frac{\partial H}{\partial p_i} - \mu_a \frac{\partial v^a}{\partial p_i} \right) + \frac{\partial H}{\partial p_i} \left( -\frac{\partial H}{\partial x^i} + \mu_a \frac{\partial v^a}{\partial x^i} \right),$$

or:

(3.8) $$\frac{dH}{dt} = \frac{\partial H}{\partial t} - \mu_a \left( \frac{\partial H}{\partial x^i} \frac{\partial v^a}{\partial p_i} - \frac{\partial H}{\partial p_i} \frac{\partial v^a}{\partial x^i} \right) = \frac{\partial H}{\partial t} - \mu_a \{H, v^a\},$$

into which we have introduced the conventional Poisson brackets:

$$\text{(3.9)} \qquad \{f, g\} = \frac{\partial f}{\partial x^i} \frac{\partial g}{\partial p_i} - \frac{\partial f}{\partial p_i} \frac{\partial g}{\partial x^i} .$$

One can see that equation (3.8) does, in fact, generalize the usual Hamilton equation for the balance of energy in the sense that one will get back to the usual equation when the $\mu_a$ vanish, i.e., when one starts from an action functional, and not the total virtual work functional.

**4. Generalized Poisson brackets.** – In equations (3.8), the conventional Poisson bracket appeared in a natural way. It should be noted that the Lie algebra that it defines on the smooth functions on a symplectic manifold $(M, \Omega)$ is in fact inherited from the Lie algebra $\mathfrak{X}(M)$ of vector fields on that manifold by way of the symplectic gradient operator, which is, in a sense, the symplectic dual of the exterior derivative [**8**]. Since vector fields that can be expressed as the symplectic gradient of a smooth function are dual to exact 1-forms under $\Omega$, one will find that it is possible to generalize the vector fields that are symplectic gradients to arbitrary vector fields by going from exact 1-forms to arbitrary 1-forms when they are expressed in Pfaffian normal form.

*a. Conventional formulation of Poisson brackets.* – In order to see how that works, let us start with the definition of the symplectic gradient of a function $f$ and how it defines the Poisson algebra. We shall denote the linear isomorphism of tangent spaces on $M$ with cotangent spaces by $\iota : T(M) \to T^*(M)$, which will take a tangent vector $\mathbf{X} \in T_x M$ to the covector $i_\mathbf{X} \Omega \in T_x^* M$, and in so doing associate a vector field $\mathbf{X}(x) \in \mathfrak{X}(M)$ with a 1-form $i_\mathbf{X} \Omega \in \Lambda^1(M)$. By abuse of notation, we shall use the symbol $\iota$ to also represent the latter invertible linear map $\iota : \mathfrak{X}(M) \to \Lambda^1(M)$, so its inverse will take a 1-form to a vector field.

If $f$ is a smooth function then $df$ will be a 1-form, and the inverse isomorphism $\iota^{-1}$ will take $df$ to a vector field:

$$\text{(4.1)} \qquad \nabla_\Omega f \equiv \iota^{-1} df$$

that we shall call the *symplectic gradient* of the function $f$, as opposed to the metric gradient that one learns about in vector calculus.

By definition, the vector field $\nabla_\Omega f$ will satisfy the equation:

$$\text{(4.2)} \qquad i_{\nabla_\Omega f} \Omega = df .$$

Thus, if $\nabla_\Omega f$ and $df$ are expressed locally in the forms:

(4.3) $$\nabla_\Omega f = (\nabla_\Omega f)^i \frac{\partial}{\partial x^i} + (\nabla_\Omega f)_i \frac{\partial}{\partial p_i}, \qquad df = \frac{\partial f}{\partial x^i} dx^i + \frac{\partial f}{\partial p_i} dp_i,$$

while $\Omega$ has the local form (3.2), then (4.2) will imply that:

(4.4) $$(\nabla_\Omega f)^i = \frac{\partial f}{\partial p_i}, \qquad (\nabla_\Omega f)_i = -\frac{\partial f}{\partial x^i},$$

i.e.:

(4.5) $$\nabla_\Omega f = \frac{\partial f}{\partial p_i} \frac{\partial}{\partial x^i} - \frac{\partial f}{\partial x^i} \frac{\partial}{\partial p_i}.$$

In particular, if $H$ is a Hamiltonian and $\mathbf{v}$ ($t$) is the velocity vector field of a curve in $M$ then Hamilton's equations can be expressed concisely in the form:

(4.6) $$\mathbf{v} = \nabla_\Omega H.$$

Thus, in general, any vector field on $M$ that can be expressed as a symplectic gradient is often referred to as a *Hamiltonian vector field*.

The way that symplectic gradients relate to the Poisson bracket is that one can define the Poisson bracket of two smooth functions $f$, $g$ on $M$ by:

(4.7) $$\{f, g\} = \Omega\, (\nabla_\Omega f, \nabla_\Omega g).$$

In order to verify that this definition is consistent with the usual definition of a Poisson bracket, note that:

$$\Omega(\nabla_\Omega f, \nabla_\Omega g) = i_{\nabla_\Omega f} \Omega(\nabla_\Omega g) = df(\nabla_\Omega g) = \left(\frac{\partial f}{\partial x^i} dx^i + \frac{\partial f}{\partial p_i} dp_i\right)\left(\frac{\partial g}{\partial p_j}\frac{\partial}{\partial x^j} - \frac{\partial g}{\partial x^j}\frac{\partial}{\partial p_j}\right)$$

$$= \frac{\partial f}{\partial x^i}\frac{\partial g}{\partial p_j} - \frac{\partial f}{\partial p_i}\frac{\partial g}{\partial x^j} = \{f, g\}.$$

An important special case of the Poisson brackets is defined when one looks at all brackets of the canonical coordinate functions on $M$, namely $\{x^i, p_i, i = 1, \ldots, n\}$. One will then get:

(4.8) $$\{x^i, x^j\} = 0, \qquad \{p_i, p_j\} = 0, \qquad \{x^i, p_j\} = \delta^i_j,$$

which are then referred to commonly as the *canonical commutation relations*. Note that under the symplectic gradient map, one will get the vector fields that correspond to the canonical coordinates in the form:

(4.9) $$\nabla_\Omega x^i = \frac{\partial}{\partial p_i}, \qquad \nabla_\Omega p_i = -\frac{\partial}{\partial x^i},$$

and the Lie brackets of all of them vanish, i.e., they all commute. Of course, that is because they represent the natural frame field for the canonical coordinate system, and natural frame fields are always holonomic.

The map $\nabla_\Omega: C^\infty(M) \to \mathfrak{X}(M)$, $f \mapsto \nabla_\Omega f$, is not only a linear isomorphism of infinite-dimensional vector spaces, it is also a homomorphism of infinite-dimensional Lie algebras, when $C^\infty(M)$ is given the Lie bracket that takes the form of the Poisson bracket ([1]). That is:

(4.10) $$\nabla_\Omega \{f, g\} = [\nabla_\Omega f, \nabla_\Omega g].$$

However, it is neither one-to-one nor onto.

It is not one-to-one because it has a non-trivial kernel, since there will be a non-trivial solution to the system of partial differential equations:

(4.11) $$0 = \nabla_\Omega f = \frac{\partial f}{\partial p_i} \frac{\partial}{\partial x^i} - \frac{\partial f}{\partial x^i} \frac{\partial}{\partial p_i},$$

i.e.:

$$0 = \frac{\partial f}{\partial x^i} = \frac{\partial f}{\partial p_i},$$

which consists of all the (locally) constant functions on $M$.

*b. Generalized Poisson brackets.* – The map is not onto either, i.e., not all vector fields on $M$ are symplectic gradients, because such vector fields must be symplectic-dual to exact 1-forms, and not all 1-forms are exact. However, every 1-form $\alpha$ (i.e., Pfaffian) can be put into Pfaffian normal form:

(4.12) $$\alpha = df + \mu_a \, dv^a \qquad (\alpha = 1, \ldots, p).$$

Since the map $\iota: \mathfrak{X}(M) \to \Lambda^1(M)$ is invertible, one can always find a unique vector field **a** that corresponds to the 1-form $\alpha$, and in fact, since $df$ will go to $\nabla_\Omega f$ for any $f$:

(4.13) $$\mathbf{a} = \iota^{-1}\alpha = \nabla_\Omega f + \mu_a \nabla_\Omega v^a.$$

Thus, the symplectic gradient operator allows one to define essentially a normal form for any vector field on $M$ that is the symplectic dual to the normal form of a 1-form.

If one defines a second vector field:

---

([1]) Indeed, when one peruses the literature that Lie produced [**9**], one will see that his definition of the Lie bracket was essentially a generalization of the Poisson bracket.

(4.14) $$\mathbf{b} = \nabla_\Omega g + \rho_A \nabla_\Omega \sigma^A \qquad (A = 1, \ldots, p')$$

then one can form the Lie bracket of the two and get:

(4.15) $$[\mathbf{a}, \mathbf{b}] = [\nabla_\Omega f + \mu_a \nabla_\Omega v^a, \nabla_\Omega g + \rho_A \nabla_\Omega \sigma^A].$$

From the bilinearity of the Lie bracket, one has:

$$[\mathbf{a}, \mathbf{b}] = [\nabla_\Omega f, \nabla_\Omega g] + [\nabla_\Omega f, \rho_A \nabla_\Omega \sigma^A] + [\mu_a \nabla_\Omega v^a, \nabla_\Omega g] + [\mu_a \nabla_\Omega v^a, \rho_A \nabla_\Omega \sigma^A].$$

From the fact that if $\lambda$ is a smooth function, while $\mathbf{X}$ and $\mathbf{Y}$ are vector fields, one has:

(4.16) $$[\mathbf{X}, \lambda \mathbf{Y}] = (\mathbf{X}\lambda)\mathbf{Y} + \lambda [\mathbf{X}, \mathbf{Y}], \qquad [\lambda \mathbf{X}, \mathbf{Y}] = \lambda [\mathbf{X}, \mathbf{Y}] - (\mathbf{Y}\lambda)\mathbf{X},$$

one can expand (4.15) into:

$$\begin{aligned}
[\mathbf{a}, \mathbf{b}] &= [\nabla_\Omega f, \nabla_\Omega g] + (\nabla_\Omega f \rho_A) \nabla_\Omega \sigma^A + \rho_A [\nabla_\Omega f, \nabla_\Omega \sigma^A] \\
&\quad - (\nabla_\Omega g \mu_a) \nabla_\Omega v^a + \mu_a [\nabla_\Omega v^a, \nabla_\Omega g] - \rho_A (\nabla_\Omega \sigma^A \mu_a) \nabla_\Omega v^a \\
&\quad + \mu_a (\nabla_\Omega v^a \rho_A) \nabla_\Omega \sigma^A + \mu_a \rho_A [\nabla_\Omega v^a, \nabla_\Omega \sigma^A],
\end{aligned}$$

and when one substitutes the Poisson brackets for the Lie brackets (when applicable), one will get:

$$\begin{aligned}
[\mathbf{a}, \mathbf{b}] &= \nabla_\Omega\{f, g\} + (\nabla_\Omega f \rho_A) \nabla_\Omega \sigma^A + \rho_A \nabla_\Omega\{f, \sigma^A\} \\
&\quad - (\nabla_\Omega g \mu_a) \nabla_\Omega v^a - \rho_A (\nabla_\Omega \sigma^A \mu_a) \nabla_\Omega v^a \\
&\quad + \mu_a \nabla_\Omega\{v^a, g\} + \mu_a (\nabla_\Omega v^a \rho_A) \nabla_\Omega \sigma^A + \mu_a \rho_A \nabla_\Omega\{v^a, \sigma^A\}.
\end{aligned}$$

With a little rearranging that will give:

(4.17) $$\begin{aligned}
[\mathbf{a}, \mathbf{b}] &= \nabla_\Omega\{f, g\} + \rho_A \nabla_\Omega\{f, \sigma^A\} + \mu_a \nabla_\Omega\{v^a, g\} + \mu_a \rho_A \nabla_\Omega\{v^a, \sigma^A\} \\
&\quad - ((\nabla_\Omega g)\mu_a + \rho_A (\nabla_\Omega \sigma^A \mu_a)) \nabla_\Omega v^a + ((\nabla_\Omega f)\rho_A + \mu_a (\nabla_\Omega v^a \rho_A)) \nabla_\Omega \sigma^A.
\end{aligned}$$

Clearly, this will reduce to the usual Poisson bracket when **a** and **b** are symplectic gradients (i.e., when $\mu_a$, $v^a$, $\rho_A$, $\sigma^A$ all vanish). One imagines that the fully-general case is probably quite intractable, but that is because one is dealing with an infinite-dimensional Lie algebra on a manifold of unspecified topology. Perhaps the elementary case in which $M$ is simply the $2n$-dimensional symplectic vector space might be more tractable.

**5. Generalized Hamilton-Jacobi equation.** – First, we shall summarize the essential points of the usual way of obtaining the Hamilton-Jacobi equation and then propose a natural generalization of it to the case in which a Hamiltonian exists only as part of an inexact 1-form.

*a. Conventional approach*. – Ordinarily, the way that one gets the Hamilton-Jacobi equation is to first represent the 1-form that defines the integrand of the action functional in Hamiltonian form, which is the *Poincaré-Cartan 1-form:*

(5.1) $$\theta = p_i \, dx^i - H \, dt.$$

That is a 1-form on the phase space $\mathcal{P}(\mathbb{R}, M)$ of the configuration manifold $M$, which has local coordinate charts that look like $(t, x, p)$. Thus, one can imagine that it projects onto the product manifold $\mathbb{R} \times M$ in the obvious way that takes $(t, x, p)$ to $(t, x)$. One then defines a section of that projection $p : \mathbb{R} \times M \to \mathcal{P}(\mathbb{R}, M)$, $(t, x) \mapsto (t, x^i, p_i(t, x))$ that one calls a *contact field*, because it associates a *contact element*, in the form of the linear hyperplane in each $T_xM$ that is annihilated by the 1-form:

(5.2) $$p = p_i(t, x) \, dx^i$$

for each value of $(t, x)$. An essential difference between a contact field of this form and the momentum 1-form $p_i(t) \, dx^i$ along a curve is that the latter is defined only along that curve, while the former must be defined everywhere in $\mathbb{R} \times M$. In effect, one is dealing with the difference between a single curve and a congruence (i.e., foliation) of curves that fills up all of $\mathbb{R} \times M$. Thus, it would not be correct to regard the current picture as equivalent to the previous Lagrangian or Hamiltonian ones, in which one considered only curve segments, but not curve congruences.

When one has such a contact field $p$, one can pull back the 1-form on $\mathcal{P}(\mathbb{R}, M)$ to a 1-form on $\mathbb{R} \times M$, which will take the local form:

(5.3) $$p^*\theta = p_i(t, x) \, dx^i - H(t, x, p(t, x)) \, dt.$$

Thus, the action functional on curves in $M$ will become a functional on curves in $\mathbb{R} \times M$:

(5.4) $$S[t, x(t)] = \int_{(t, x(t))} p^*\theta = \int_{(t, x(t))} [p_i(t, x(t)) \, dx^i - H(t, x, p(t, x(t))) \, dt],$$

and if $\dot{x}^i(t) = dx^i / dt$ then this can also be written in the form:

$$(5.5) \qquad S[t, x(t)] = \int_{(t, x(t))} [p_i(t, x(t)) \dot{x}^i - H(t, x, p(t, x(t)))] \, dt.$$

In general, that functional will be path-dependent. However, the restriction that will produce the Hamilton-Jacobi equation is that it is not. In effect, the action *functional* $S[t, x(t)]$ will then reduce to an action *function* $S[t_0, x_0; t_1, x_1]$, in which $(t_0, x_0)$ and $(t_1, x_1)$ are the initial and final points of the curve segment $(t, x(t))$, respectively. That will be possible if and only if the 1-form $p^*\theta$ is exact, i.e., there must be a smooth function $S(t, x)$ on $\mathbb{R} \times M$ such that:

$$(5.6) \qquad p^*\theta = dS.$$

If one expands the right-hand side into:

$$(5.7) \qquad dS = \frac{\partial S}{\partial t} dt + \frac{\partial S}{\partial x^i} dx^i$$

and compares that to (5.3) then that will give the pair of equations:

$$(5.8) \qquad p_i(t, x) = \frac{\partial S}{\partial x^i}, \qquad -H(t, x, p(t, x)) = \frac{\partial S}{\partial t}.$$

The first one implies that the contact field $p$ must be spatially-closed:

$$(5.9) \qquad 0 = d_{s\wedge} p = \tfrac{1}{2}(\partial_i p_j - \partial_j p_i) \, dx^i \wedge dx^j.$$

Thus, not every contact field can be used to pull back the Poincaré-Cartan 1-form to $\mathbb{R} \times M$. In effect, such a contact field must have vanishing "vorticity," i.e., it must be "irrotational." More to the point, the exterior differential system (i.e., Pfaffian system) $p^*\theta = 0$ on $\mathbb{R} \times M$ must be completely integrable into a codimension-one foliation of level hypersurfaces of $S(t, x)$.

When one substitutes the first equation in (5.8) into the second one, one can express the latter in the form:

$$(5.10) \qquad \frac{\partial S}{\partial t} + H\left(t, x^i, \frac{\partial S}{\partial x^i}\right) = 0,$$

which is then the *Hamilton-Jacobi equation* in its conventional form. It is therefore a first-order partial differential equation for the function $S$ whose linearity will depend upon the nature of $H$.

Since most of the Hamiltonians that occur in physics tend to be quadratic in the components of the momentum 1-form, that means that typically the partial differential equation will be nonlinear.

*b. Generalization of the Hamilton-Jacobi equation.* – If one no longer has merely a Hamiltonian *H*, but only a 1-form that replaces *dH*, namely:

(5.11) $$\eta = dH - \mu_a \, dv^a,$$

then the Poincaré-Cartan 1-form cannot be generalized directly, but only its exterior derivative:

(5.12) $$d_\wedge \theta = dp_i \wedge dx^i - dH \wedge dt.$$

A natural generalization is to *replace dH* with $\eta$ and define the 2-form on $\mathcal{P}(\mathbb{R}, M)$ :

(5.13) $$\Theta = dp_i \wedge dx^i - \eta \wedge dt = d_\wedge \theta - d\mu_a \wedge dv^a.$$

(Note that a simple *substitution* of *dH* with $\eta + \mu_a \, dv^a$ would not change anything fundamentally.)

If one has a contact field *p* on $\mathbb{R} \times M$, as before, then one can pull that 2-form down to a 2-form $p^*\Theta$ on $\mathbb{R} \times M$, which will then amount to:

(5.14) $$p^*\Theta = (p^*dp_i) \wedge dx^i - p^*\eta \wedge dt.$$

Since:

$$p^*dp_i = \frac{\partial p_i}{\partial t} dt + \frac{\partial p_i}{\partial x^j} dx^j,$$

so

$$(p^*dp_i) \wedge dx^i = \frac{\partial p_i}{\partial t} dt \wedge dx^i + \tfrac{1}{2}(\partial_i p_j - \partial_j p_i) dx^i \wedge dx^j,$$

and

$$p^*\eta = p^*dH - \mu_a p^*dv^a = \left(\frac{\partial H}{\partial t} - \mu_a \frac{\partial v^a}{\partial t}\right) dt + \left(\frac{\partial H}{\partial x^i} - \mu_a \frac{\partial v^a}{\partial x^i}\right) dx^i,$$

which makes:

$$-p^*\eta \wedge dt = -\left(\frac{\partial H}{\partial x^i} - \mu_a \frac{\partial v^a}{\partial x^i}\right) dx^i \wedge dt = \left(\frac{\partial H}{\partial x^i} - \mu_a \frac{\partial v^a}{\partial x^i}\right) dt \wedge dx^i,$$

one will have:

$$(5.15) \qquad p^*\Theta = \left(\frac{\partial p_i}{\partial t} + \frac{\partial H}{\partial x^i} - \mu_a \frac{\partial v^a}{\partial x^i}\right) dt \wedge dx^i + \tfrac{1}{2}(\partial_i p_j - \partial_j p_i) dx^i \wedge dx^j .$$

Instead of generalizing the exactness of $p^*\theta$, which would now be meaningless, we shall generalize the idea that if $p^*\theta$ is exact then it must be closed, i.e., $d_\wedge p^*\theta = 0$. However, since exterior differentiation commutes with pull-backs, that will be equivalent to demanding that $p^* d_\wedge \theta = 0$. Since we are replacing $d_\wedge \theta$ with $\Theta$, the natural generalization would be:

$$(5.16) \qquad p^*\Theta = 0 ,$$

and based upon (5.15), that will give the pair of equations:

$$(5.17) \qquad \frac{\partial p_i}{\partial t} + \frac{\partial H}{\partial x^i} - \mu_a \frac{\partial v^a}{\partial x^i} = 0 , \qquad \partial_i p_j - \partial_j p_i = 0 .$$

That system now constitutes a system of $n$ first-order (generally nonlinear) partial differential equations for the contact field $p$. The second one still says that the contact field must be spatially irrotational. Hence, it is once more spatially exact, so $p_i = \partial S / \partial x^i$, and the first system of equations will take the form:

$$(5.18) \qquad \frac{\partial}{\partial x^i}\left(\frac{\partial S}{\partial t} + H\right) = \mu_a \frac{\partial v^a}{\partial x^i} .$$

When the 1-form $\eta$ reduces to $dH$ (so the $\mu_a$ and $v^a$ all vanish), the latter equation will say that the expression in parentheses on the left-hand side must be a function of only time, say $\xi(t)$, which will give:

$$(5.19) \qquad \frac{\partial S}{\partial t} + H = \xi(t) ,$$

although $\xi(t)$ can be absorbed into the definition of $H$, so that is still basically the Hamilton-Jacobi equation.

**6. Discussion.** – For this author, the actual driving ambition behind the generalization of the principle of least action was to find a better way of discussing the difference between classical mechanics and quantum wave mechanics at the fundamental level. Indeed, a recurring theme in quantum theory seems to be the idea that quantum physics equals classical physics plus quantum fluctuations. For instance, in the path integral approach to quantum mechanics, one must consider not only classical extremals, such as geodesics, but also neighboring non-extremal paths, in order to arrive at the correct quantum expressions for the relativistic scattering amplitudes (i.e., transition probabilities). Similarly, quantum conservation laws, such as the Ward-Takahashi identities, appear to take the form of replacing the zero that would appear on the right-hand side of the classical conservation law with non-zero quantum corrections. The loop expansions that appear in effective field theories of quantum phenomena also seem to embody that notion that quantum equals classical plus corrections that typically arise from the existence of vacuum polarization and the interaction of the field of a "particle" with that quantum vacuum.

If one returns to the optical mechanical analogy that motivated de Broglie and Schrödinger to formulate the earliest form of wave mechanics then one sees that the difference between classical and quantum mechanics is, presumably, analogous to the difference between geometrical optics and wave optics. Now, geometrical optics is typically subordinate to a severe limitation, namely, that one deals with waves of such short wavelength that diffraction effects can be ignored. In a sense, this also the approximation that one must consider only a single light ray and its infinitesimal neighborhood, whereas wave optics considers congruences of light rays with a finite cross section. Thus, one might also say that wave optics equals geometrical optics plus diffraction corrections. Indeed, the asymptotic expansions that play a key role in the extension of geometrical optics by diffraction corrections, namely, WKB expansions, are closely analogous to the loop expansions of effective quantum field theories.

Since the common element in both geometrical optics and classical mechanics would seem to be that of a geodesic, one might reasonably suspect that a promising direction of progress for the generalization of classical mechanics might be that of considering geodesics plus some sort of quantum correction. For instance, one might return to the Lagrange equations in their first form, rather than their second (which is, as we said, the Euler-Lagrange equations) and look into the possible form that the non-conservative force 1-form on the right-hand side might take if it is to represent quantum corrections, in some sense of the term. That would be reasonable insofar as the Lagrangian for geodesics typically takes the form of a sort of "kinetic energy" term. The problem is then to find a physically-meaningful replacement for the zero on the right-hand side of the geodesic equation that will probably take the form of an inexact 1-form. One suspects that since radiation is a dissipative process the force of radiation damping might be represented in that way, although perhaps not in its Lorentz-Dirac formulation, which is based on point-like charges (i.e., single geodesics, not congruences of them). The author has made a first attempt at formulating equations of geodesics with quantum correction in [**10**].

———————